\documentclass[fleqn,twoside]{article}
\usepackage{amssymb}
\usepackage{espcrc2}

\usepackage{graphicx}
\usepackage[figuresright]{rotating}

\newcommand{\kcrit}{\mbox{$\kappa_{\rm crit}$}}

\newcommand{\ksea}{\mbox{$\kappa_{\rm sea}$}}
\newcommand{\kval}{\mbox{$\kappa_{\rm val}$}}

\newcommand{\AmS}{{\protect\the\textfont2
  A\kern-.1667em\lower.5ex\hbox{M}\kern-.125emS}}

\title{Light quark masses from UKQCD's dynamical simulations with O(a)-improved Wilson fermions }

\author{Derek J. Hepburn\address[MCSD]{School of Physics, University of Edinburgh, Edinburgh EH9 3JZ, UK.}, UKQCD Collaboration.}
       
\begin{document}

\begin{abstract}
I present preliminary results on the light quark masses from a partially
quenched analysis of UKQCD's dynamical datasets.
\vspace{1pc}
\end{abstract}

\maketitle

\section{Introduction}

The determination of the masses of the light quarks represents an
important challenge in particle physics today. They are among the
least well determined parameters of the Standard Model, having
uncertainties of the same order as their values as given in the
particle data book.

Lattice QCD is in a position to improve this situation because it
allows one to study the quark mass dependence of physical observables
such as hadron masses. The physical values of these observables can
then be used to fix the quark masses. Unfortunately current
simulations cannot probe the up/down quark mass range directly and so
some form of extrapolation is necessary. Chiral perturbation theory is
at present the best procedure for providing a link to the low mass
regime. The simulations studied in this paper however are still too
heavy for $\chi^{PT}$ and a simpler extrapolation ansatz is used instead.

In this paper I present results for the average up/down quark mass,
$m_l$, and the strange quark mass, $m_s$, from a partially quenched
analysis of pseudoscalar and vector meson masses for each of UKQCD's
dynamical fermion datasets. The suitably averaged values of the hadron
masses used to fix the quark masses are as follows: $M_\pi=137.3\
\textrm{MeV}$ ; $M_K=495.7\ \textrm{MeV}$ ; $M_{K^*}=893.9\
\textrm{MeV}$ and $M_\phi=1020\ \textrm{MeV}$.

\section{Simulation Details}

All of the lattice simulations studied here have been carried out on a
$16^3\times32$ lattice with a non-perturbatively O(a)-improved Wilson
fermion action and standard Wilson glue. Details of these simulations
have been discussed previously in
\cite{csw202,mylatt2001}. Additionally the set of $\kval$ values for
the datasets at $\beta=5.2, \ksea=0.13565$ and $\beta=5.2,
\ksea=0.1358$ have been increased to $\{0.1342, 0.1350, 0.1355,
0.13565, 0.13580\}$. Pseudoscalar and vector meson correlation
functions have been generated with all non-degenerate combinations of
these $\kval$ values. For the $\ksea=0.1358$ dataset these correlators
have also been calculated at $\kval=0.13595$.

\section{Hadron masses}

The procedures used here to extract hadron masses from correlation
functions have been described in detail in \cite{csw202,mylatt2001}.
In brief the masses of the pseudoscalar and vector mesons were
determined from double cosh fits to their corresponding correlation
functions. Local and fuzzed correlators were fitted simultaneously to
constrain fit parameters and a sliding window analysis used to find an
appropriate time window for each fit.

\section{Chiral Extrapolations}

The pseudoscalar meson masses for each dataset were extrapolated in
$1/\kval$ using the simple ansatz

\begin{eqnarray}
M_{PS}^2(\kappa_1,\kappa_2) & = & A_{PS} ( m_1 + m_2 )\\
& = & A_{PS} (\ \frac{1}{2}(\frac{1}{\kappa_1} + \frac{1}{\kappa_2}) - \frac{1}{\kcrit} )
\end{eqnarray}
The value of $\kcrit$ was determined as a free parameter in this fit.

For the vector meson the ansatz
\begin{equation}
M_V(m_1,m_2) = A_V ( m_1 + m_2 ) + B_V
\end{equation}
was found to fit the data well. 

\begin{figure}[ht]
\includegraphics*[width=7.5cm,height=7.5cm]{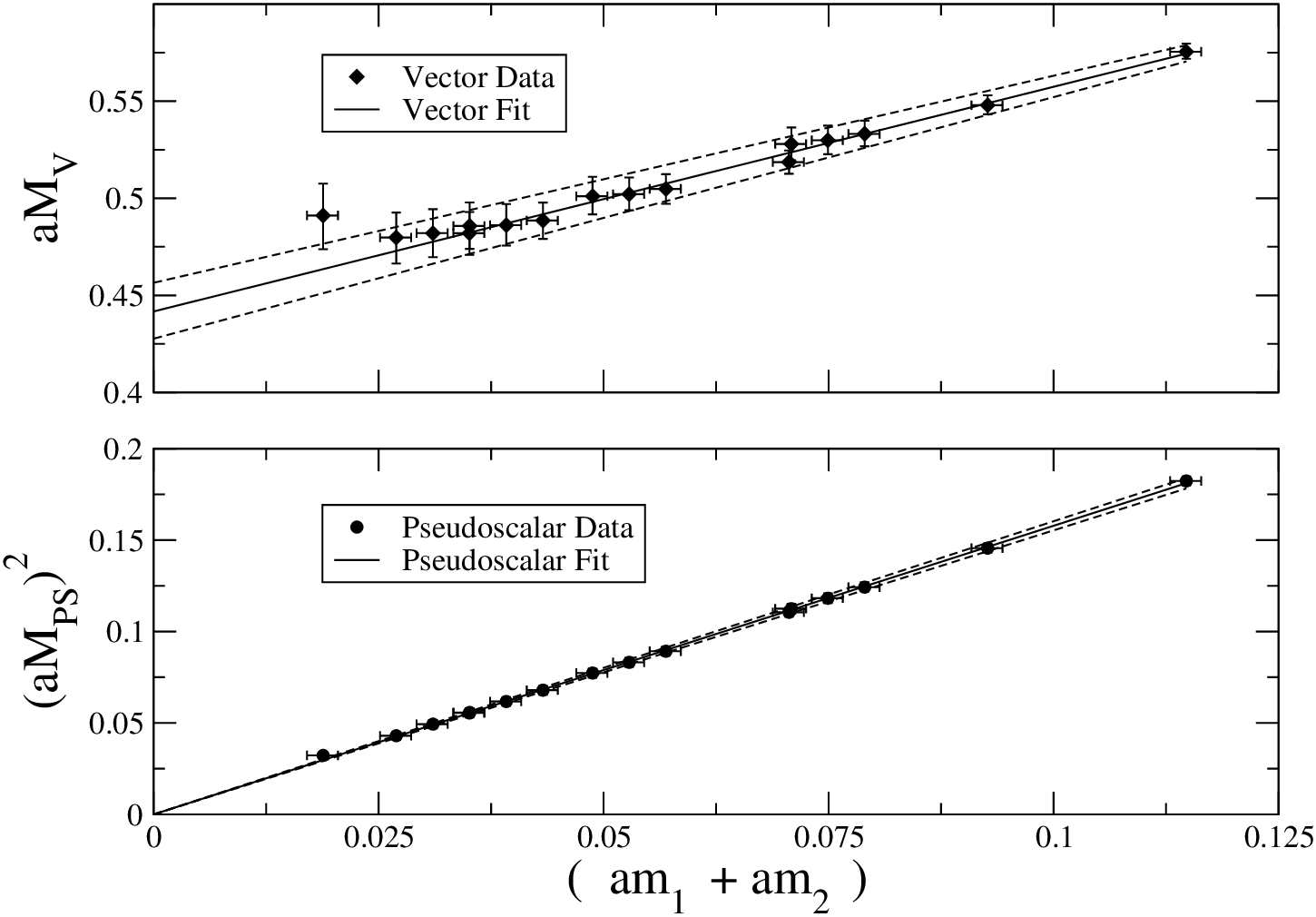}
\vspace{-1.0cm}
\caption{Chiral extrapolations of pseudoscalar and vector meson masses
for the $\beta=5.2, \ksea=0.1358$ dataset.}
\vspace{-0.5cm}
\label{extrap}
\end{figure}

An example plot of these chiral extrapolations is given in Figure
\ref{extrap} for the $\ksea=0.1358$ dataset. The lightest pseudoscalar
and vector points at $\kval=0.13595$ are included in the plot but were
not included in the extrapolation. The lightest vector point clearly
deviates from the linear behaviour suggesting large finite size
effects. For more discussion of finite size effects in these datasets
I refer the reader to \cite{lightalan}.

\section{Fixing the bare light quark masses}

Determination of the light quark masses from the results of the chiral
extrapolations requires setting the scale by some physical
quantity. In this work the Sommer scale parameter,
$r_0=0.49\ \textrm{fm}$, and $M_\rho=769.3\ \textrm{MeV}$ were used.

The average up/down quark mass, $m_l$, was determined by requiring
that the pseudoscalar mass equal the physical pion mass at that point i.e.

\begin{equation}
   M_{PS}^2(m_l,m_l) = a M_\pi
\end{equation}

The strange quark mass, $m_s$, was then determined from each of $M_K$,
$M_{K^*}$ and $M_\phi$ by solving

\begin{equation}
  M_{PS}^2(m_l,m_s) = a M_K
\end{equation}
\begin{equation}
  M_{V}(m_l,m_s) = a M_{K^*}
\end{equation}
\begin{equation}
  M_{V}(m_s,m_s) = a M_\phi
\end{equation}

When the lattice spacing was fixed by $r_0$ however only the $M_K$
method gave a sensible value for $m_s$ and only these results are
given.

\section{Renormalisation and matching}

In this work the values for the renormalised quark masses are given in
the $\overline{MS}$ scheme at a reference scale of $2\ \textrm{GeV}$. The
relationship between the renormalised and bare quark masses is given
by the relation

\begin{equation}
  m_R^{\overline{MS}}(\mu) = Z_m^{\overline{MS}}(\mu)(1 + b_m am_q)m_q
\end{equation}
where the matching has been performed at a scale, $\mu$. Running the
masses from the scale, $\mu$, to a scale of $2\ \textrm{GeV}$ was achieved
using the Mathematica package, RunDec \cite{rundec}, which implements
4-loop running from perturbation theory.

One-loop tadpole-improved perturbative results \cite{alpha_bm,Zm} were
used for $Z_m$ and $b_m$ reorganised in terms of
$\alpha^{\overline{MS}}(\mu)$

\begin{equation}
  Z_m^{\overline{MS}}(\mu) = [ 1 -
  \frac{\alpha^{\overline{MS}}(\mu)}{4\pi} ( 8\ \ln(a\mu) - 15.085 ) ]u_0^{-1}
\end{equation}
\begin{equation}
  b_m = [ -1/2 - 0.7850\ \alpha^{\overline{MS}}(\mu) ] u_0^{-1}
\end{equation}
The tadpole factor, $u_0$, was taken to be $1/8\kappa_c$ here.

An estimation of the best value for the matching scale, $\mu=q^*$
\cite{lepage}, has not been determined for this work. Instead the
cases $\mu=1/a$ and $\mu=\pi/a$ have been investigated and the
variation treated as a systematic effect.

The values of $\alpha^{\overline{MS}}(\mu)$ were also determined with
the RunDec package using values for $\Lambda_{\overline{MS}}$ given in
\cite{lambda}.

\section{Results}

\begin{figure}[ht]
\includegraphics*[width=7.5cm,height=6.5cm]{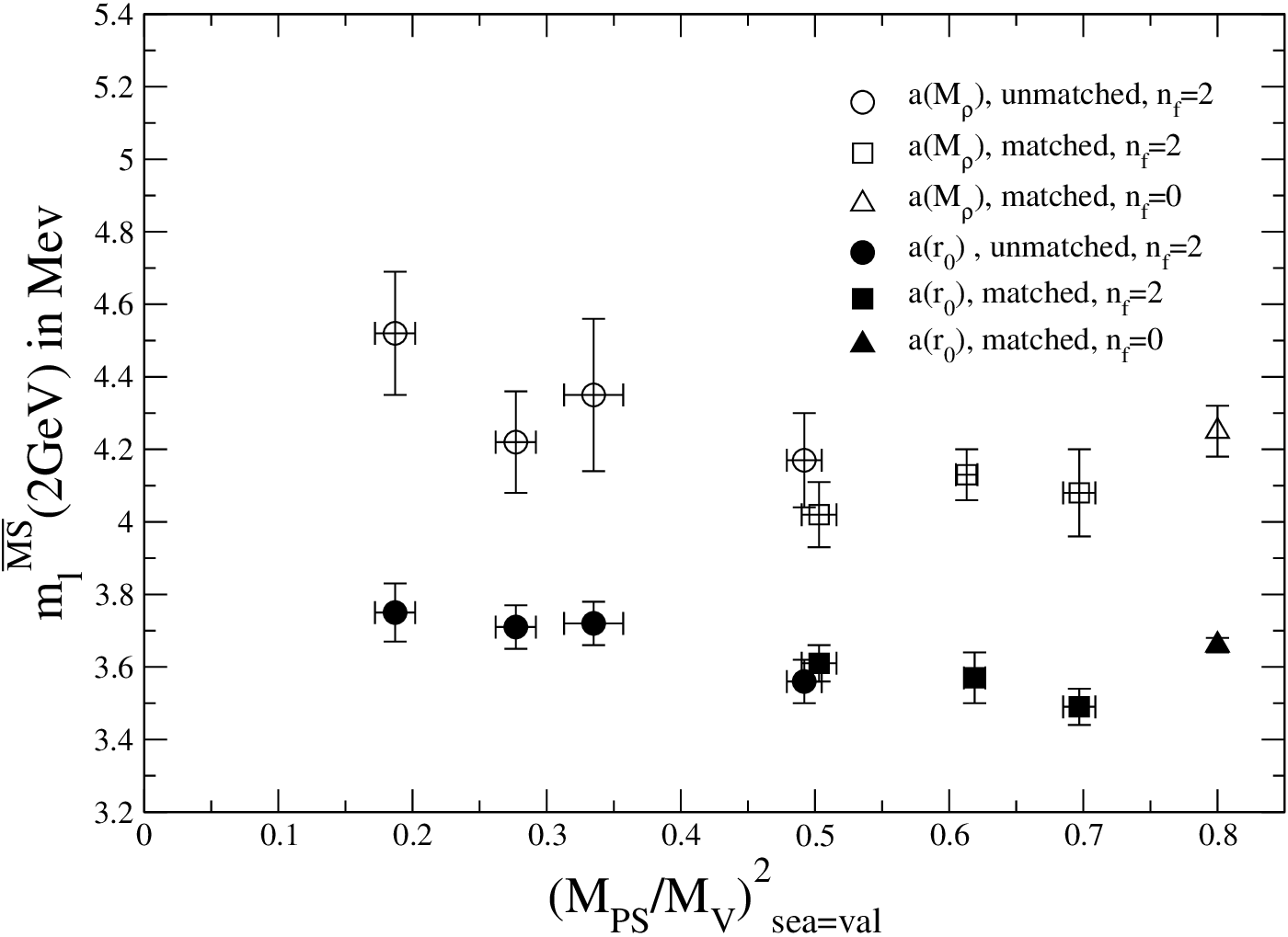}
\vspace{-1.0cm}
\caption{Average up/down quark masses with $\mu=1/a$. Only statistical errors are shown.}
\vspace{-0.5cm}
\label{mlight}
\end{figure}

\begin{figure}[ht]
\includegraphics*[width=7.5cm,height=6.5cm]{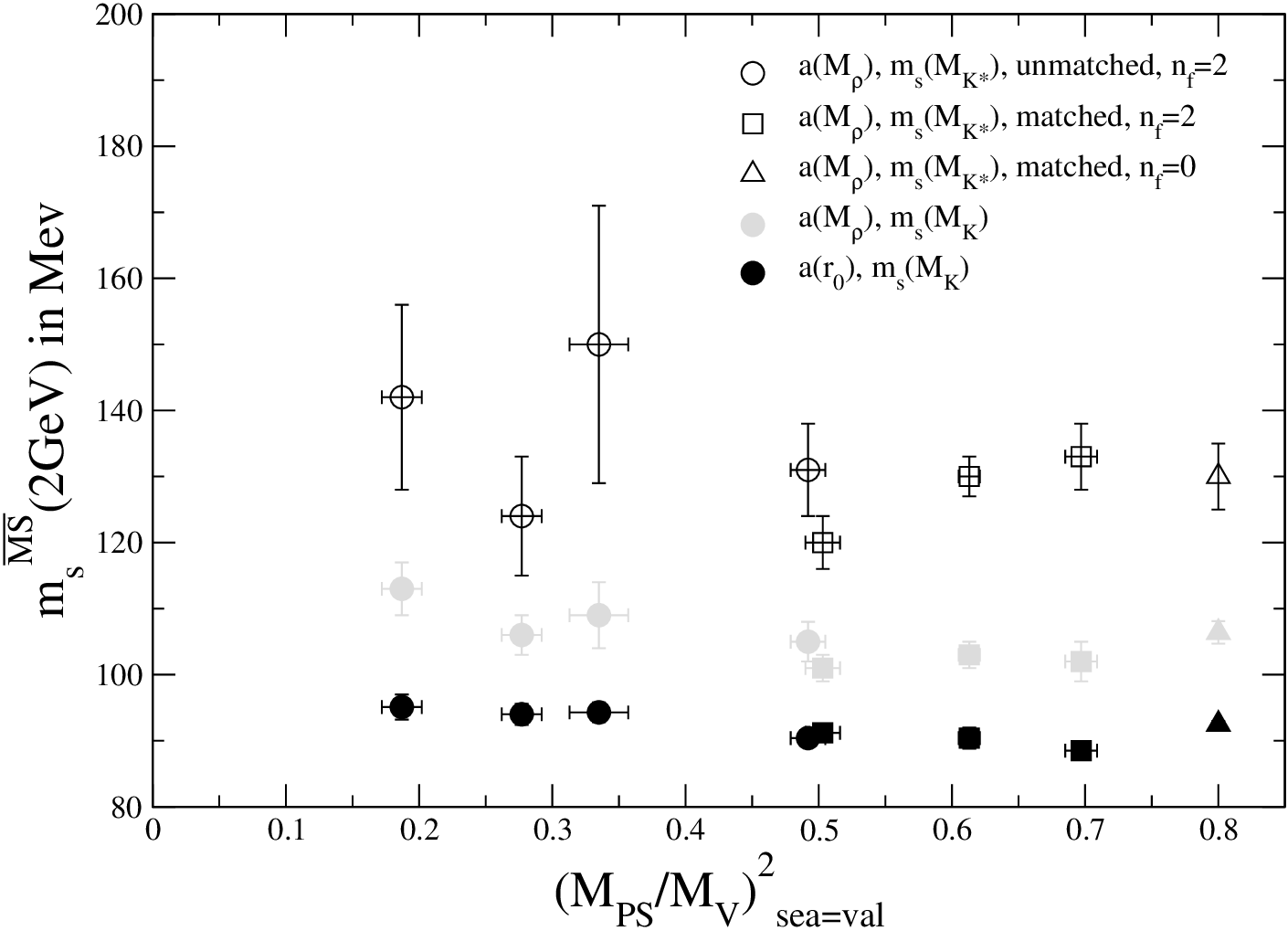}
\vspace{-1.0cm}
\caption{Strange quark masses with $\mu=1/a$. Only statistical errors are shown.}
\vspace{-0.5cm}
\label{mstrange}
\end{figure}

Results for $m_l$ and $m_s$ have been plotted in Figures \ref{mlight}
and \ref{mstrange} respectively. Datasets matched in $r_0$ have been
distinguished from the others and a matched quenched dataset included
for comparison. Only results for $\mu=1/a$ are plotted as choice of
matching scale was found to have around a $1\textrm{\%}$ effect. The
results for $m_s(M_\phi)$ are omitted as they were almost identical to
those for $m_s(M_{K^*})$.

A large dependence on the choice of scale setting parameter can be
seen for both $m_l$ and $m_s$. It should be noted however that a
smaller physical value for $r_0$ would bring the results into better
agreement. There is also a clear difference between $m_s(M_K)$ and
$m_s(M_{K^*})$ though this would be expected to vanish in the
continuum limit.

There is some indication of a trend in the matched datasets as the sea
quark mass is decreased, as expected. When compared to the matched
quenched point however there is no clear signal of unquenching from
either quark mass though this may improve with statistics.

The ratio, $m_s/m_l$, was calculated for each scheme and found to be
between $25.0$ and $25.5$ for $m_s(M_K)$ and around $31.0$ for
$m_s(M_{K^*})$. This can be compared with the result from $\chi^{PT}$
\cite{leutwyler} of $24.4(1.5)$.

\section{Acknowledgments}

I would like to thank the Carnegie Trust for the Universities of
Scotland, the European Community's Human Potential programme under
HPRN-CT-2000-00145 Hadrons/LatticeQCD and PPARC under grant
PP/G/S/1998/00777 for their financial support of this work. I also
thank Craig McNeile for his invaluable help.


\begin{thebibliography}{99}
\bibitem{csw202} C.R.Allton {\it et al.}, Phys.Rev. D65 (2002) 054502, hep-lat/0107021.
\bibitem{mylatt2001} D.J.Hepburn, Nucl.Phys.Proc.Suppl. 106
(2002) 278-280, hep-lat/0110029.
\bibitem{lightalan} A.C.Irving, this conference. 
\bibitem{rundec} K.G.Chetyrkin {\it et al.}, Comput. Phys. Commun. 133
(2000) 43-65, hep-ph/0004189.
\bibitem{alpha_bm} S.Sint and P.Weisz, Nucl.Phys. B502(1997) 251-268,
hep-lat/9704001.
\bibitem{Zm} S.Capitani {\it et al.}, Nucl.Phys.Proc.Suppl. 63 (1998)
874-876, hep-lat/9709049.
\bibitem{lepage} G.P.Lepage and P.B.Mackenzie, Phys.Rev. D48 (1993)
2250-2264, hep-lat/9209022.
\bibitem{lambda} S.Booth {\it et al.}, Phys.Lett. B519 (2001) 229-237,
hep-lat/0103023.
\bibitem{leutwyler} H.Leutwyler, Phys.Lett. B378 (1996) 313-318.
\end{thebibliography}
\end{document}